\documentclass[usenatbib,usegraphicx]{mn2e}
\usepackage{amssymb}
\usepackage{bm}
\usepackage{color}
\usepackage{graphicx}
\usepackage[normalem]{ulem}

\newcommand{\halfspace}{\hspace{1pt}}

\newcommand{\Lya}{Ly$\alpha$}
\newcommand{\Lyb}{Ly$\beta$}
\newcommand\HI{{\hbox{H\halfspace$\rm \scriptstyle I$}}}

\newcommand\HeII{{\hbox{He\halfspace$\rm \scriptstyle II$}}}

\def\bkt#1{\left(#1\right)} 
\def\bkts#1{\left[#1\right]} 
\def\bkta#1{\langle#1\rangle} 
\newcommand\gsim{~\lower.5ex\hbox{$\buildrel > \over \sim$}~}
\newcommand\lsim{~\lower.5ex\hbox{$\buildrel < \over \sim$}~}
\def\mb#1{\mathbf#1}
\def\mc#1{\mathcal{#1}}

\title[UVBG fluctuations and the LyA forest]{The influence of metagalactic ultra-violet background fluctuations on the  high-redshift Ly$\alpha$\ forest}

\author[Avery Meiksin]{
        Avery Meiksin$^{1}$\thanks{E-mail:\ A.Meiksin@ed.ac.uk}\\
        $^{1}$SUPA\thanks{Scottish Universities Physics Alliance},
	Institute for Astronomy, University of Edinburgh,
        Blackford Hill, Edinburgh\ EH9\ 3HJ, UK}

\shortcites{}

\begin{document}

\date{Accepted 29 November 2019. Received 13 November 2019; in original form 25 September 2019 }
\pagerange{\pageref{firstpage}--\pageref{lastpage}} \pubyear{2019}
\maketitle
\label{firstpage}

\begin{abstract}
 Under the assumption that galaxies and Quasi-Stellar Objects (QSOs) dominate the metagalactic ultra-violet (UV) background, it is shown that at high redshifts fluctuations in the UV background are dominated by QSO shot noise and have an auto-correlation length of a few to several comoving Mpcs, depending on the bright end of the QSO luminosity function. The correlations create long range spatial coherence in the neutral hydrogen fraction. Using a semi-analytic model, it is demonstrated that the coherence may account for the broad distribution in effective optical depths measured in the \Lya\ forest spectra of background QSOs, for line-of-sight segments of comoving length $50h^{-1}$~Mpc at redshifts $5<z<6$. Capturing the fluctuations in a numerical simulation requires a comoving box size of $\sim1h^{-1}$~Gpc, although a box half this size may be adequate if sufficient random realizations of the QSO population are performed.
\end{abstract}


\section{Introduction}
\label{sec:Intro}

Establishing the Epoch of Reionization (EoR), when most of the baryons in the Universe were re-ionized following the recombination era, is a major goal of observational programmes over a wide range of wavebands. Measurements of the Cosmic Microwave Background suggest reionization largely occurred in the redshift interval $5<z<10$, with a characteristic redshift for the EoR of $z_{\rm reion}= 7.6\pm0.7$ \citep{2018arXiv180706209P}. The \Lya\ emission line profiles of high redshift Quasi-Stellar Objects (QSOs) and  \Lya-emitting galaxies suggest reionization was still underway at $z\sim7$ \citep{2018Natur.553..473B, 2018ApJ...856....2M}. The search for a 21-cm signature from the Intergalactic Medium (IGM) during the EoR is a primary driver of a new generation of radio interferometers \citep{2013ExA....36..235M}, with preliminary results beginning to arrive \citep{2016ApJ...833..102B, 2019MNRAS.488.4271G}.

The sources that reionized the Universe are unknown, but are widely expected to be dominated by early galaxies, with a smaller contribution from QSOs \citep[e.g.][]{2019ApJ...879...36F, 2019MNRAS.485...47P}. The mass range of the galaxies providing most of the photoionizing radiation is under contention because of the uncertain star-formation histories and ultra-violet (UV) spectra of high redshift galaxies \citep{2009ApJ...690.1350O, 2015ApJ...811..140B, 2017ApJ...835..113L}, and the uncertain escape fractions of the ionizing photons \citep{2013AA...553A.106L, 2006ApJ...643...75N, 2015MNRAS.451.2544P, 2017MNRAS.468.4077A}. Whilst it has been suggested low mass galaxies are the primary drivers of reionization by virtue of a stellar population with a high ionization efficiency or a high escape fraction \citep[eg][]{2015ApJ...802L..19R, 2017MNRAS.468.4077A, 2017ApJ...835..113L}, models in which massive galaxies dominate have been suggested as well \citep{2019arXiv190713130N}. A possibility remains that more exotic sources, such as decaying dark matter particles \citep[eg][]{2016JCAP...08..054O} or cosmic strings \citep{2019arXiv190708022L}, may also have contributed to or even dominated the reionization.

Following the discovery of high redshift QSOs in large numbers by the Sloan Digital Sky Survey \citep{2006AJ....131.2766R, 2018AA...613A..51P}, it has become possible to probe large-scale fluctuations in the \Lya\ forest into the EoR. Early results showed a rapid rise in the mean absorption of the IGM towards increasing redshift, with increasing scatter \citep{2006AJ....132..117F}. Enhancing the data set continues to show wide variance in the optical depths, with troughs resembling the Gunn-Peterson effect \citep{1965ApJ...142.1633G}, extending over comoving lines of sight with lengths up to $\sim100h^{-1}\,{\rm Mpc}$, suggesting reionization may still be ongoing at $z\sim6$ \citep{2015MNRAS.447.3402B}.

Attempts to model the fluctuations within conventional models of the IGM and the UV background, whilst matching the distributions of optical depths at $z\lsim5$, are unable to reproduce the widths of the distributions at higher redshifts without invoking additional assumptions regarding the structure of the IGM not well substantiated by direct measurements, such as large IGM temperature fluctuations following reionization \citep{2015ApJ...813L..38D}, a much shorter ionizing photon mean free path than indicated by QSO spectra \citep{2016MNRAS.460.1328D, 2018ApJ...863...92B, 2018MNRAS.473..560D} or reionization ending late, persisting until $z\lsim6$ \citep{2015MNRAS.447.3402B, 2018MNRAS.479.1055B}. The recovery of the optical depth distributions using reionization simulations, with reionization completing as late as $z\sim5.2$, supports the latter possibility \citep{2019MNRAS.485L..24K, 2019MNRAS.tmp.2682K}.

Another solution invokes a large contribution of QSOs, or other rare, highly luminous sources, to the UV photoionizing background at $z>5$, comparable to the galactic contribution. Such a scenario has been advocated by \citet{2015A&A...578A..83G, 2019ApJ...884...19G} based on the high numbers of QSOs they find compared with previous surveys. Simulations find that the UV background fluctuations produced by a boosted QSO population result in a broadened effective optical depth distribution for the \Lya\ forest at $z>5$, in good agreement with measurements \citep{2015MNRAS.453.2943C, 2017MNRAS.465.3429C}.

In this paper, an alternative explanation is examined. Fluctuations in the UV background depend on the variance in the luminosity of the sources, both for their 1-point distribution \citep{1992MNRAS.258...36Z} and 2-point spatial correlations \citep{1992MNRAS.258...45Z}. Whilst QSOs provide only a small fraction of the mean ionizing intensity at $z>5$, they dominate the fluctuations. It is shown here that source shot noise from both QSOs and galaxies produces large-scale fluctuations in the UV background and a consequent large-scale coherence in the \Lya\ forest absorption, on the scale of several comoving Mpcs, with results sensitive to the numbers of high luminosity QSOs. Very large simulation volumes, with comoving box sizes on the order of $\sim1h^{-1}\,{\rm Gpc}$, are required to adequately capture the fluctuations. The broad \Lya\ forest optical depth distributions at $5<z<6$ are expected for some standard QSO luminosity functions. Indeed, the optical depth distribution may be informing us as much about the high luminosity tail of the QSO luminosity function as about cosmic reionization.

This paper is structured as follows:\ the modelling assumptions are described in the next section. The results are presented in Sec.3, followed by a Discussion in Sec.4 and a summary of the main Conclusions in Sec.5. An Appendix provides technical details on the modelling. All numerical results assume cosmological parameter values for a $\Lambda$CDM cosmology consistent with {\it Planck} 2018 measurements  \citep{2018arXiv180706209P}. The notation \lq cMpc\rq\ refers to comoving Mpc.
  
\section{Large-scale fluctuations in the \Lya\ forest}
\label{sec:bgflucs}

\subsection{UV background model}
\label{subsec:uvbgmodel}

The mean UV background and its fluctuations are modelled following \citet{2019MNRAS.482.4777M}. The comoving emissivity has the form
\begin{equation}
\epsilon_\nu(z)=\epsilon_L\left(\frac{\nu}{\nu_L}\right)^{-\alpha_j}(1+z)^{-\alpha_S},
\label{eq:emiss}
\end{equation}
where $\epsilon_L$ is a normalization factor, and $\nu_L$ denotes the threshold frequency for photoelectric absorption. The angle-averaged intensity is given by
\begin{equation}
4\pi J_\nu(z) =
\int_z^\infty\,dz'\frac{dl_p}{dz'}\epsilon_{\nu'}(z')(1+z)^3e^{-\tau_\nu(z,z')},
\label{eq:Jnu}
\end{equation}
where $dl_p/dz = c/[H(z)(1+z)]$, $\epsilon_{\nu'}$ is the comoving
emissivity, $\nu'=\nu(1+z')/(1+z)$, and $\tau_\nu(z,z')$ is the optical depth due to IGM attenuation along a path from $z'$ to $z$. Details of the attenuation model are provided in \citet{2019MNRAS.482.4777M}. The hydrogen photo-ionization rate is
\begin{equation}
\Gamma_{\rm H}(z) = \int_{\nu_L}^\infty d\nu\,\frac{4\pi J_\nu}{h_P\nu}\sigma_\nu,
\label{eq:Gamma}
\end{equation}
where $\sigma_\nu$ is the photoelectric cross-section. A good match to UV background estimates over redshifts $2<z<6$ \citep{2012ApJ...746..125H, 2019MNRAS.485...47P} is provided by the choices $\alpha_j=1.8$ and $\alpha_S=0.8$.

The emissivity is modelled as having two contributions, arising from
QSO and galaxy sources. The QSO component uses the results of
\citet{2007ApJ...654..731H}. Three of their luminosity function models
are considered here:\ the full redshift evolution fit to a
double-power law luminosity function ($z$-ev), the pure luminosity
evolution fit (PLE) and a modified Schechter function fit (mS). They
span the behaviour of the high luminosity end of the QSO luminosity
function, essential for quantifying the UV background
fluctuations. The contributions of the models to the UV background at the Lyman edge are estimated following \citet{2007ApJ...654..731H}. Unless stated otherwise, the full redshift evolution model is used for the computations, as it provides the best-fitting and most complete description of the QSO data
\citep{2007ApJ...654..731H}. For the smaller density-dependent
contribution to the UV background fluctuations, an evolving QSO bias
factor $b_Q=0.278(1+z)^2+0.57$ is adopted, based on results from the
extended-BOSS QSO survey \citep{2017JCAP...07..017L}.

Two other QSO luminosity functions will be referred to for comparison, from \citet{2019MNRAS.488.1035K} and \citet{2019ApJ...884...19G}, who provide luminosity functions at restframe wavelength $1450$A. Adopting the assumed spectra in the papers, the contributions of the QSOs to the UV background are based on a QSO spectral shape $f_\nu\sim\nu^{-0.61}$ for \citet{2019MNRAS.488.1035K} and $f_\nu\sim\nu^{-0.44}$ over $1200\,{\rm A}<\lambda<1450$A and $f_\nu\sim\nu^{-1.57}$ for $\lambda<1200$A for \citet{2019ApJ...884...19G}.

The galaxy luminosity function used is from
\citet{2015ApJ...803...34B}, along with an ionizing photon escape
fraction of $f_{\rm esc}=1.8\times10^{-4}(1+z)^{3.4}$
\citep{2012ApJ...746..125H}. A galaxy bias factor of $b_G=3$ is used
\citep[e.g.][]{2013MNRAS.430..425B}, although it may be substantially
higher at $z>5$. The galaxies dilute the shot noise component of the
UV background power spectrum compared with the QSO-only case. At
$z=3$, the contributions of the galaxies and QSOs to the mean UV
background are comparable, but the QSO contribution decreases to
$\sim10$ percent at $z>5$. The QSOs, however, dominate the
fluctuations in the UV background over all redshifts considered here.

\begin{figure}
\scalebox{0.43}{\includegraphics{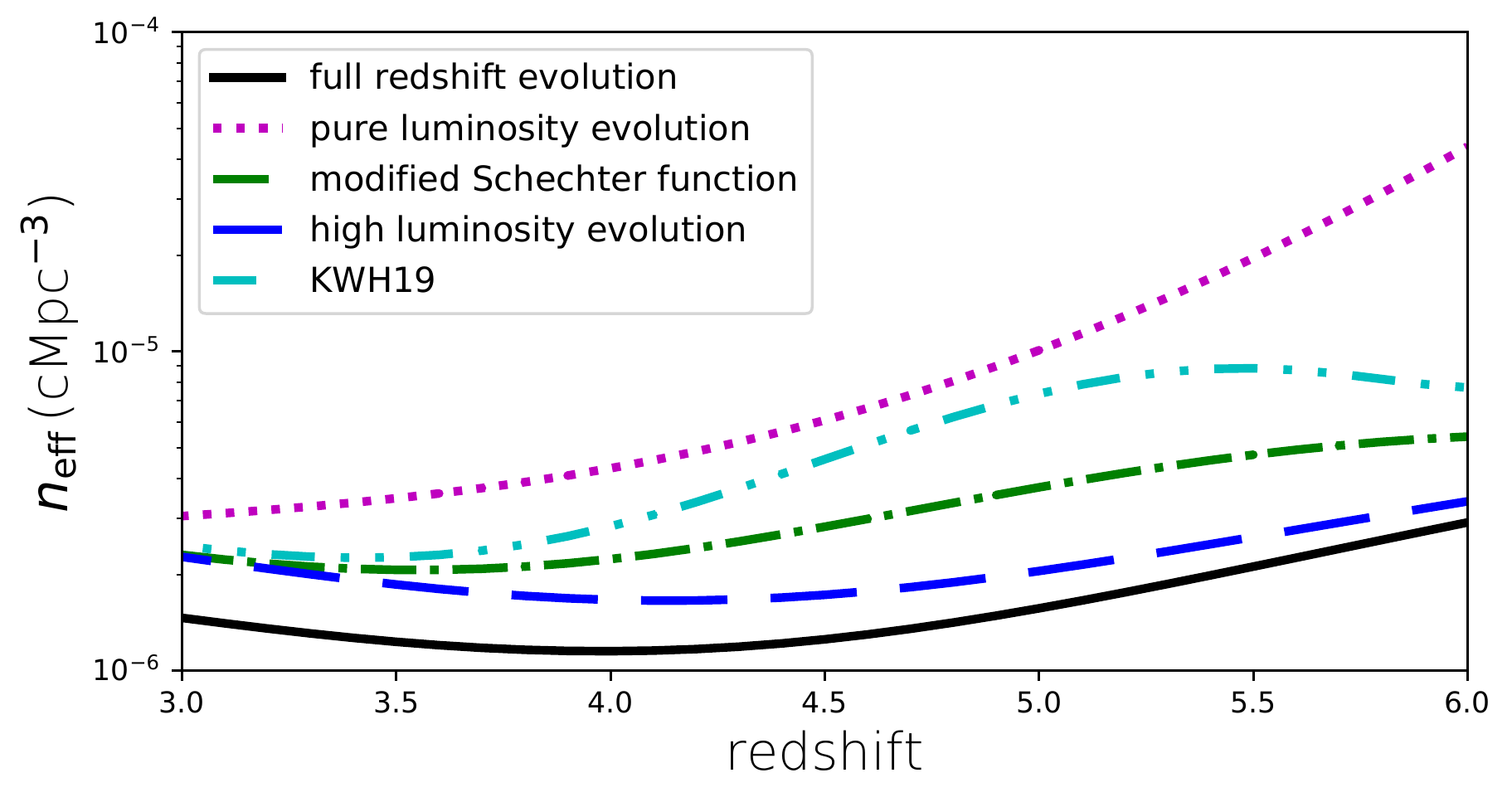}}
\caption{Effective (comoving) number density $n_{\rm eff}$ of sources including both galaxies and QSOs. Results are shown for four different QSO luminosity functions from \citet{2007ApJ...654..731H}:\ a full redshift evolution model (black solid line), a pure luminosity evolution model (magenta dotted line), a modified Schechter luminosity function (green dot-dashed line), and a redshift evolution fit to the high luminosity end of the luminosity function (blue dashed line). Also shown is the result for Model 3 of \citet{2019MNRAS.488.1035K} (cyan dot-dot-dashed line).}
\label{fig:neff}
\end{figure}

The variance in the luminosity of sources may be characterized by the effective mean number density of sources

\begin{equation}
  n_{\rm eff} = \frac{\left[\int_0^\infty\,dL\,L\Phi(L)\right]^2}{\int_0^\infty\,dL\,L^2\Phi(L)},
  \label{eq:neff}
\end{equation}
where $\Phi(L)$ is the luminosity density function of sources of luminosity $L$. The shot noise contribution to the spatial correlations in the UV background is proportional to $n_{\rm eff}^{-1}$. Even allowing for the high number density of galaxies, the fluctuations are very sensitive to the QSO luminosity function, as illustrated in Fig.~\ref{fig:neff}, which shows the effective number density at the Lyman edge for four QSO luminosity function models of \citet{2007ApJ...654..731H}. Only QSOs with bolometric luminosities between $10^{10}<L/L_\odot<10^{15}$ are considered. At high luminosities, the luminosity function varies as $L^{-\beta}$ with $\beta<3$. This corresponds to $\langle L^2\rangle\sim L_{\rm max}^{3-\beta}$, increasing with the maximum QSO bolometric lumunosity $L_{\rm max}$, and exceeding the galaxy contribution to $\langle L^2\rangle$. In addition to the three models listed above, the effective density is also shown for a fourth model from \citet{2007ApJ...654..731H}, for which a redshift evolution model is fit to the high luminosity tail of the luminosity function. It roughly interpolates between the full redshift evolution model and the modified Schechter function model at these redshifts, and so is not further considered.

Alternative QSO double-power-law luminosity functions have been proposed by \citet{2019MNRAS.488.1035K}, which have a much steeper luminosity dependence for high luminosities compared with the models of \citet{2007ApJ...654..731H}. The models, however, require a more rapidly increasing break luminosity with redshift. The resulting effective number density for their Model 3, integrated over $-30<M_{1450}<-21$, is included in Fig.~\ref{fig:neff}. (The effective densities for Models 1 and 2 are similar at the redshifts shown.) Since it lies between the modified Schechter function and pure luminosity evolution models at the redshifts of interest, it is not further pursued here. It is noted that the low luminosity power-law exponent is smaller than 3, and so the effective number density is sensitive to the break luminosity.

  Very similar values for the effective number density are found using the luminosity function of \citet{2019ApJ...884...19G}, although the galaxy contribution must be partly suppressed so as not exceed estimates for the total metagalactic emissivity, by about 30 percent. The reduction in the galaxy contribution has the effect of reducing the effective number density of sources to values comparable to other predictions. For example, at $z=5.6$, the full redshift evolution model of  \citet{2007ApJ...654..731H} gives $n_{\rm eff}\simeq2\times10^{-6}\,{\rm cMpc}^{-3}$. The luminosity function of \citet{2019MNRAS.488.1035K} gives $n_{\rm eff}\simeq9\times10^{-6}\,{\rm cMpc}^{-3}$, while \citet{2019ApJ...884...19G} gives $n_{\rm eff}\simeq6\times10^{-6}\,{\rm cMpc}^{-3}$ (allowing for a $\sim30$ percent reduction in the galaxy contribution to the metagalactic emissivity). Adopting the same spectral shape as used in \citet{2019MNRAS.488.1035K} gives instead $n_{\rm eff}\simeq4\times10^{-6}\,{\rm cMpc}^{-3}$.

\subsection{\Lya\ forest model}
\label{subsec:LyAmodel}

A full description of the impact of UV background fluctuations on the \Lya\ forest requires large-scale coupled hydrodynamical-gravity simulations with radiative transfer. Such simulations are computationally very expensive, limiting the parameter ranges that may be searched. An inexpensive approximate alternative is to model the \Lya\ forest using dark matter only. A simplified version is adopted here based on the log-normal approximation for the dark matter density field \citep{1997ApJ...479..523B}. As only absorption properties averaged over velocity scales broad compared with the absorption features are considered, the method should provide an adequate description to estimate the impact of the UV background fluctuations on the properties of interest. The formalism is developed following \citet{2014MNRAS.437.3639C}.

The log-normal model approximates the baryon number density $n_b$  for mild overdensities according to
\begin{equation}
  n_b(\mb x , z)=n_0(z) \exp\bkts{\delta_b (\mb   x,z) - \bkta{\delta_b^2(\mb x, z)}/2},
  \label{eq:logn}
\end{equation}
where $n_0(z)$ is the mean baryon density and $\delta_b(\mb x, z)=n_b(\mb x, z)/n_0(z)-1$ the baryon density fluctuation at (comoving) position $\mb x$ and redshift $z$. The baryon density fluctuations are derived from the dark-matter density perturbations by Jeans-filtering the dark-matter fluctuations, with filter $W(k) = 1/[1 + (x_Jk)^2]$, where $x_J$ is the comoving Jeans length, given by
\begin{equation}
  x_J = H_0^{-1}\bkt{2\gamma k_{\rm B} T_m(z)\over 3\mu m_p\Omega_m(1+z)}^{1/2},
\end{equation}
for a present Hubble constant $H_0$, matter density parameter $\Omega_m$ and $k_{\rm B}$ is the Boltzmann constant. Here $T_m(z)$ is the density-averaged IGM temperature, $\mu=4/(8-5Y)$ is the mean molecular weight of the IGM for helium mass abundance $Y$, and $\gamma$ is the polytropic index defined by the IGM equation of state
\begin{equation}
  T(\mb x, z)= T_0(z) \bkt{n_b(\mb x, z)\over n_0(z)}^{\gamma(z) -1},
  \label{eq:EoS}
  \end{equation}
where $T_0(z)$ is the temperature at mean density, to which $T_m(z)$ is set. The redshift dependent parameters $T_0(z)$ and $\gamma(z)$ are adopted from \citet{2011MNRAS.410.1096B}. The results are not very sensitive to these values. From the continuity equation, the peculiar velocity perturbation corresponding to a baryon perturbation $\delta_b({\mb k}, t)$ is given by ${\bm v}({\bm k}, t) = {\bm u}({\bm k}, t)/ (1+z)$, where
\begin{equation}
{\mb u}({\mb k}, t) = -i\frac{\dot D(t)}{D(t)}\hat{\mb k}w(k, t),
\end{equation}
with $\hat{\mb k}={\mb k}/k$ for $k=\vert{\mb k}\vert$, $w(k, t)=\delta_b(k, t)/k$ and $D(t)$ is the linear density fluctuation growth factor.

To construct the line-of-sight \Lya\ forest spectrum, only the line-of-sight peculiar velocity is required. This may be constructed from the joint power spectrum between the density field and line-of-sight velocity component, which introduces two correlated random variables. Because the UV background arises from sources that follow the dark matter density field, the model must be extended to allow also for statistical correlations between all three of the density, line-of-sight velocity and UV background fields. The details are provided in an Appendix.

The shot noise contribution to the UV background fluctuations must be treated separately, as these fluctuations are uncorrelated with the underyling density field. Because the shot noise contribution to the power spectrum varies as $k^{-2}$ at large $k$ \citep{2019MNRAS.482.4777M}, the corresponding 1D projected power spectrum formally diverges. (The variance of the 1-point distribution similarly diverges.) This is primarily a technical issue, in so far that the \Lya\ forest signature is a filtered representation of the UV background fluctuations. (Physically, the amplitude of the fluctuations are limited on small scales by the sizes of the emitting systems.) The shot noise contribution in the model is accordingly regulated by filtering. To ensure the spatial correlations of the shot noise contribution are recovered, it is necessary to preserve the non-linear character of the fluctuations. A log-normal representation is found adequate, although other approximations are considered in the Appendix. Accordingly, the hydrogen photoionization rate is given by
\begin{eqnarray}
  \Gamma_{\rm H}({\mb x}, z) &=& \Gamma_{{\rm H}, 0}(z)\left[1+\delta_{\Gamma, b}({\mb x}, z)\right]\nonumber\\
&&\times  \exp\bkts{\delta_{\Gamma, {\rm sn}}({\mb x}, z) - \bkta{\delta_{\Gamma, {\rm sn}}^2({\mb x}, z)}/2},
  \label{eq:Gammalogn}
\end{eqnarray}
where $\Gamma_{{\rm H}, 0}(z)$ is the mean photoionization rate at redshift $z$, $\delta_{\Gamma, b}$ represents fluctuations correlated with the density field and $\delta_{\Gamma, {\rm sn}}$ represents the shot-noise contribution, filtered on a scale for which $\bkta{\delta_{\Gamma, {\rm sn}}^2({\mb x}, z)}$ is of order unity. It is shown in the Appendix that the results are not very sensitive to the choice of filter scale as long as the scale is short compared with the scale on which the fluctuations are measured. Both the density-dependent and shot noise contributions to the power spectrum of the photoionizing background fluctuations are computed numerically following \citet{2019MNRAS.482.4777M}.

A line of sight simulated must be sufficiently long to capture the modes required to reproduce the UV background spatial correlations. In the Appendix, it is shown that a comoving line of sight length of $400h^{-1}\,{\rm cMpc}$ is adequate. For the results presented here, lines of sight of length $800h^{-1}\,{\rm cMpc}$ are used. To ensure complete statistical independence in the effective optical depths averaged over segments of width $\Delta x$, only a single random segment is used to compute the effective optical depth per random line of sight. The effective optical depth distributions are based on 4096 independent random realizations.

\section{Results}
\label{sec:results}

\subsection{UV background spatial correlations}
\label{subsec:UV backgroundfluc}

\begin{figure}
\scalebox{0.55}{\includegraphics{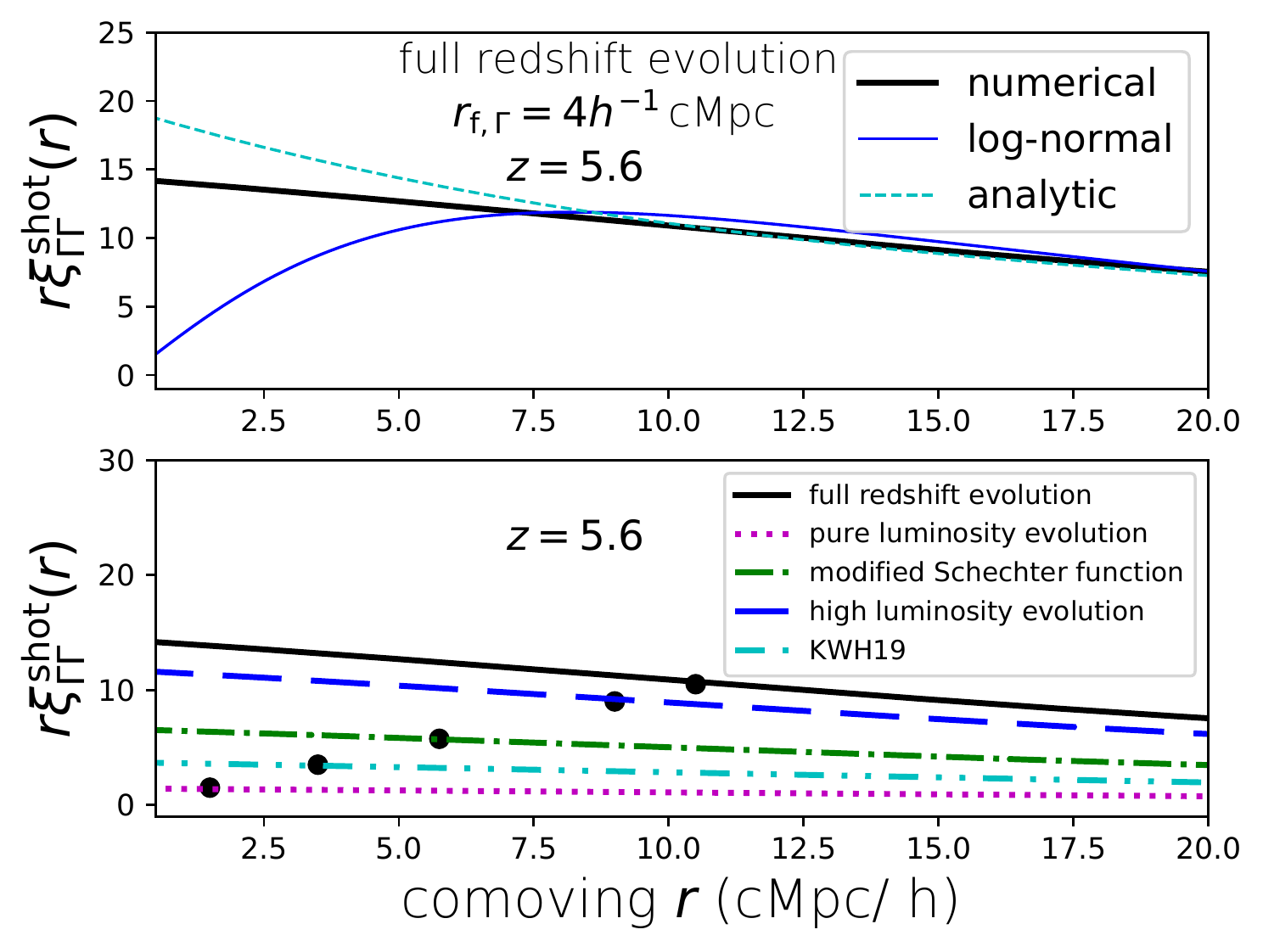}}
\caption{Shot-noise component of the spatial auto-correlation function of the UV background, $r\xi_{\Gamma\Gamma}^{\rm shot}(r)$, as a function of comoving spatial separation $r$ ($h^{-1}\,$cMpc), at $z=5.6$. (Top panel)\ The three curves show the numerical solution (heavy black solid line), the matching log-normal approximation for a smoothing length $r_{{\rm f}, \Gamma}=4h^{-1}\,{\rm cMpc}$ (thin blue solid line), and the analytic solution (cyan dashed line; see text). Results are shown including both galaxy and QSO sources, assuming the full redshift evolution model for the QSO luminosity function. (Bottom panel)\ The spatial auto-correlation function for five QSO luminosity function models, as indicated. The scale of the correlation lengths, where $\xi_{\Gamma\Gamma}^{\rm shot}=1$, is indicated by a black dot for each model.}
\label{fig:rxiGG}
\end{figure}

The low effective number density of sources produces large-scale spatial correlations in the UV background, as illustrated by the full redshift evolution model at $z=5.6$ in Fig.~\ref{fig:rxiGG} (top panel), where $\xi^{\rm shot}_{\Gamma \Gamma}(r)=\langle\delta_{\Gamma, {\rm sn}}(0)\delta_{\Gamma, {\rm sn}}(r)\rangle$. Also shown is the analytic prediction \citep{1992MNRAS.258...45Z}. The result is comparable to the numerical prediction, although the numerically computed correlations are somewhat weaker on small scales. The scales are sufficiently small to have a negligible effect on the effective optical depth distribution.

In the bottom panel, the numerically computed correlation functions are shown for five QSO luminosity functions. The comoving correlation length $r_{0, \Gamma}$, defined by $\xi_{\Gamma\Gamma}^{\rm shot}(r_{0, \Gamma})=1$, is also indicated for each model. The correlation length varies from $10h^{-1}\,{\rm cMpc}$ for the full redshift evolution model to $1h^{-1}\,{\rm cMpc}$ in the PLE model. The correlation functions are shallow, declining only somewhat more steeply than $1/r$, reaching  0.1 only beyond $40h^{-1}\,{\rm cMpc}$ for the full redshift evolution model and $10h^{-1}\,{\rm cMpc}$ for the PLE model. They will produce large-scale spatial correlations in the \HI\ fraction, and so in the \Lya\ forest optical depths \citep{2019MNRAS.482.4777M}.

As shown in the top panel of Fig.~\ref{fig:rxiGG}, the log-normal model for the UV background fluctuations well reproduces the numerically computed shot-noise component of the correlation function on scales for which $\xi_{\Gamma\Gamma}^{\rm shot}>0.5$, for a filter scale $r_{\rm f, \Gamma}=4h^{-1}\,{\rm cMpc}$ for the full redshift evolution model, except below the filter scale, where the filtering suppresses the correlations. The filter scale for the other QSO luminosity functions decreases like the correlation length. For the PLE model, it is $r_{\rm f, \Gamma}=0.5h^{-1}\,{\rm cMpc}$. As shown in the Appendix, the \Lya\ effective optical depth distributions are only weakly dependent on the choice of filter scale. At larger separations, the correlation functions in the log-normal UV background model lie below the numerical computations, but as the correlations are already weak they have a negligible effect on the \Lya\ optical depth distributions.

The numerical results shown are for the steady-state solution to the evolution equation for the UV background fluctuations, which assumes the sources to be infinitely long-lived. A case for which the QSO sources have a finite lifetime is considered below.

\subsection{\Lya\ forest effective optical depth distribution}
\label{subsec:tauefffluc}

\begin{figure}
\scalebox{0.55}{\includegraphics{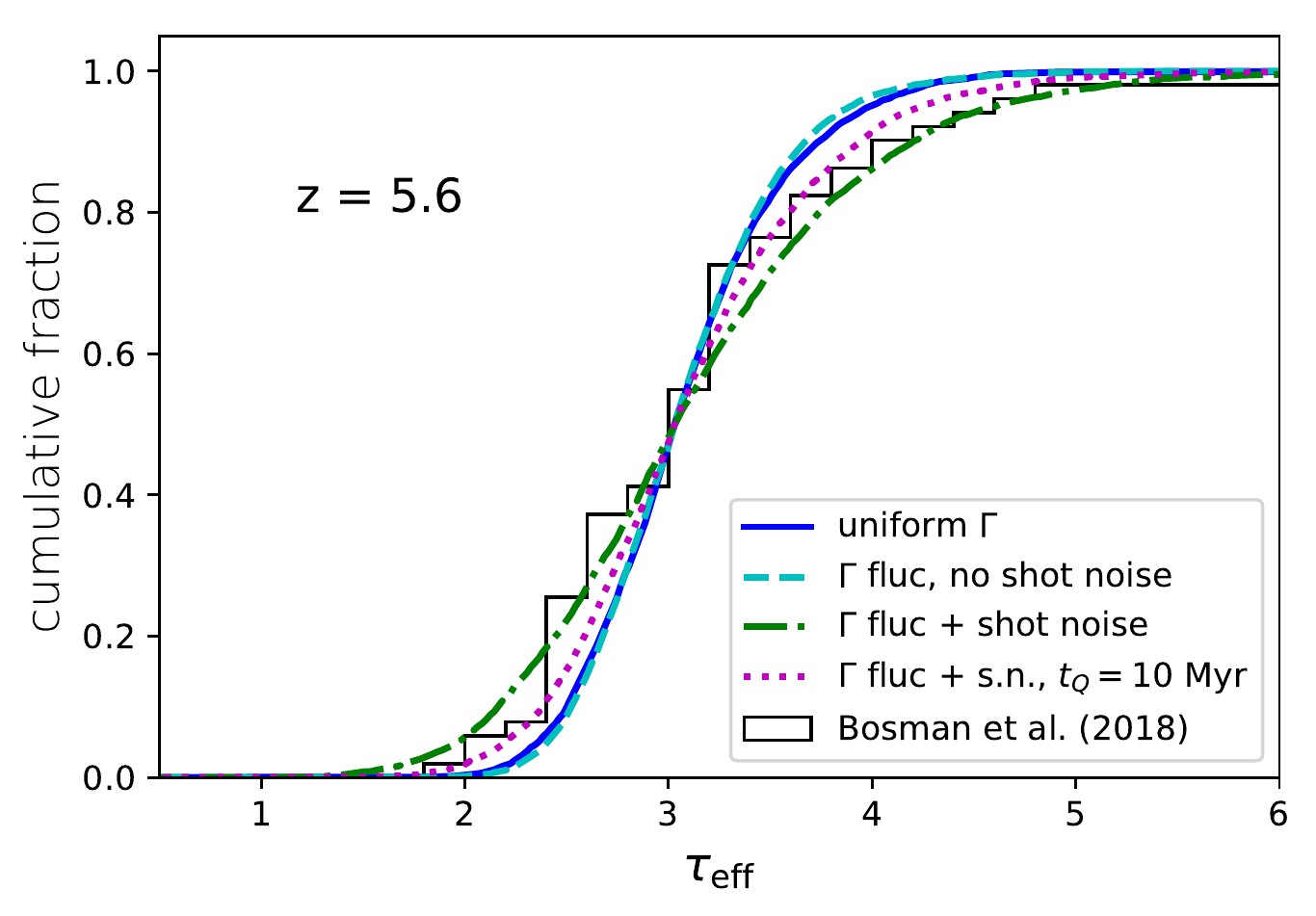}}
\caption{Effect of UV background spatial correlations on the effective optical depth distribution of the \Lya\ forest. The cumulative fractions are shown at $z=5.6$ assuming a uniform UV background (solid blue line), allowing for density-generated fluctuations in the steady-state limit, but excluding the effect of shot noise in the sources (cyan dashed line), including both density-generated and shot-noise generated UV background fluctuations, in the steady-state limit (green dot-dashed line), and including both density-generated and shot-noise generated UV background fluctuations, but assuming a QSO lifetime of 10~Myr (magenta dotted line). The histogram shows the data from \citet{2018MNRAS.479.1055B}.}
\label{fig:xiGGtaueff}
\end{figure}

The large-scale spatial correlations in the UV background produces large-scale coherence in the \Lya\ forest transmission along the lines of sight to background QSO sources. Large fluctuations result in a wide range of effective \Lya\ forest optical depths, defined by
\begin{equation}
  \tau_{\rm eff}=-\log\langle\exp(-\tau)\rangle_{\Delta x},
  \label{eq:taueff}
\end{equation}
where $\langle\exp(-\tau)\rangle$ is the mean transmission over a comoving length $\Delta x$. The effect of UV background correlations is illustrated in Fig.~\ref{fig:xiGGtaueff} at $z=5.6$. The emissivity includes galaxies and QSOs, assuming the full redshift evolution QSO luminosity function. The effective optical depths correspond to transmissions averaged over comoving segments of width $\Delta x=50h^{-1}\,{\rm cMpc}$ to compare with the measurements of \citet{2018MNRAS.479.1055B} (their SILVER sample is used). The mean UV background is normalized to match the median measured effective optical depth. In the absence of any UV background fluctuations, the optical depth distribution is considerably narrower than the measured. Allowing for UV background fluctuations arising only from density fluctuations (which will affect the mean free path of the ionizing photons), slightly tightens the distribution. This is a consequence of the expected anti-correlation between the \Lya\ absorption and density-dependent contribution to the photoionization rate fluctuation, which tends to suppress the power in the \Lya\ forest transmission \citep{2019MNRAS.482.4777M}.

By contrast, including the effects of shot noise in the sources broadens the distribution, producing a good match to the measurements. Two models are considered, UV background fluctuations in the steady state limit (dot-dashed line) and time-dependent fluctuations, allowing for a short QSO lifetime of 10~Myr (dotted line). The finite lifetime of the QSOs suppresses the shot noise power on wavelengths long compared with the photon mean free path at the Lyman edge \citep{2019MNRAS.482.4777M}. Since the UV background correlations are already weak on this scale, the time-dependent model prediction differs only slightly from that of the steady-state model.

\begin{figure*}
\includegraphics{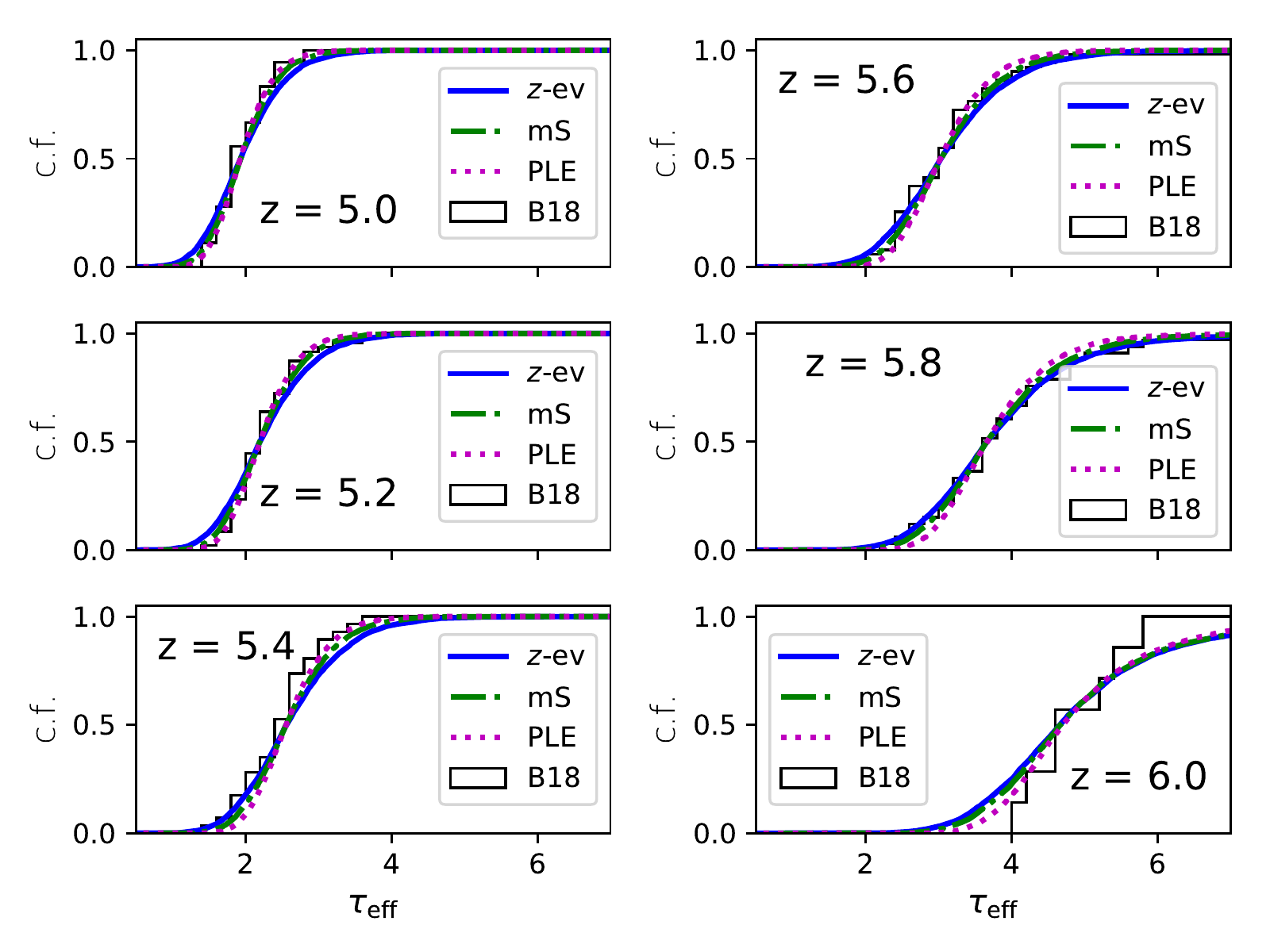}
\caption{Predicted cumulative fractions of the \Lya\ effective optical depth at redshifts $5<z<6$ resulting from fluctuations in the UV background generated by galaxies and QSOs,  allowing for three different QSO luminosity functions:\ full redshift evolution (blue solid lines), a modified Schechter function (green dot-dashed lines) and pure luminosity evolution (magenta dotted lines). The histograms show the data from \citet{2018MNRAS.479.1055B}.}
\label{fig:taueff_z5t6}
\end{figure*}

In Fig.~\ref{fig:taueff_z5t6}, the cumulative fractions over $5<z<6$ are shown for three QSO luminosity models. The modified Schechter function predictions lie between the full redshift evolution and PLE models. Generally good agreement is found for all the models for $z\leq5.5$. For $5.6<z<5.8$, the full redshift evolution and modified Schechter models are preferred over the PLE model. The PLE model matches best at $z=6$, although none are satisfactory, suggesting reionization may still be incomplete at these redshifts. Given the uncertainties in the models, in particular the upper QSO luminosities, as well as uncertainties in the data \citep[see the discussion in][of the three sample classes defined]{2018MNRAS.479.1055B}, these differences may perhaps not be statistically significant. But the models do illustrate that allowing for the shot-noise induced spatial correlations in the UV background broadens the effective optical depth distributions to a level that well matches the measured distributions at $z<6$ without invoking incomplete reionization.

The mean UV background values required to match the median optical depths are comparable to previous estimates. The required photoionization rate decreases from $\Gamma_{{\rm H}, 0}\simeq0.6\times10^{-12}\,{\rm s}^{-1}$ to $0.1\times10^{-12}\,{\rm s}^{-1}$ over $z=5$ to $z=6$. These values are typically $\sim50$ percent greater than the values for a uniform UV background, comparable to previous estimates of the boosting of the required mean photoionization rate when allowing for UV background fluctuations \citep{MW03}.

\section{Discussion}
\label{sec:discussion}

The diminishing number count of QSO sources at increasing redshift produces a large shot noise contribution to the resulting UV background fluctuations. This is a somewhat paradoxical result, as the mean UV background is increasingly dominated by galaxies at $z>4$. Because of the relative shallowness of the QSO luminosity function at the bright end, however, the QSO contribution dominates the fluctuations in the UV background, resulting in large-scale spatial correlations in the UV background. These correlations produce large-scale spatial correlations in the \Lya\ optical depth, resulting in broad distributions in the effective optical depths averaged over wide spatial segments.

\begin{figure}
\scalebox{0.43}{\includegraphics{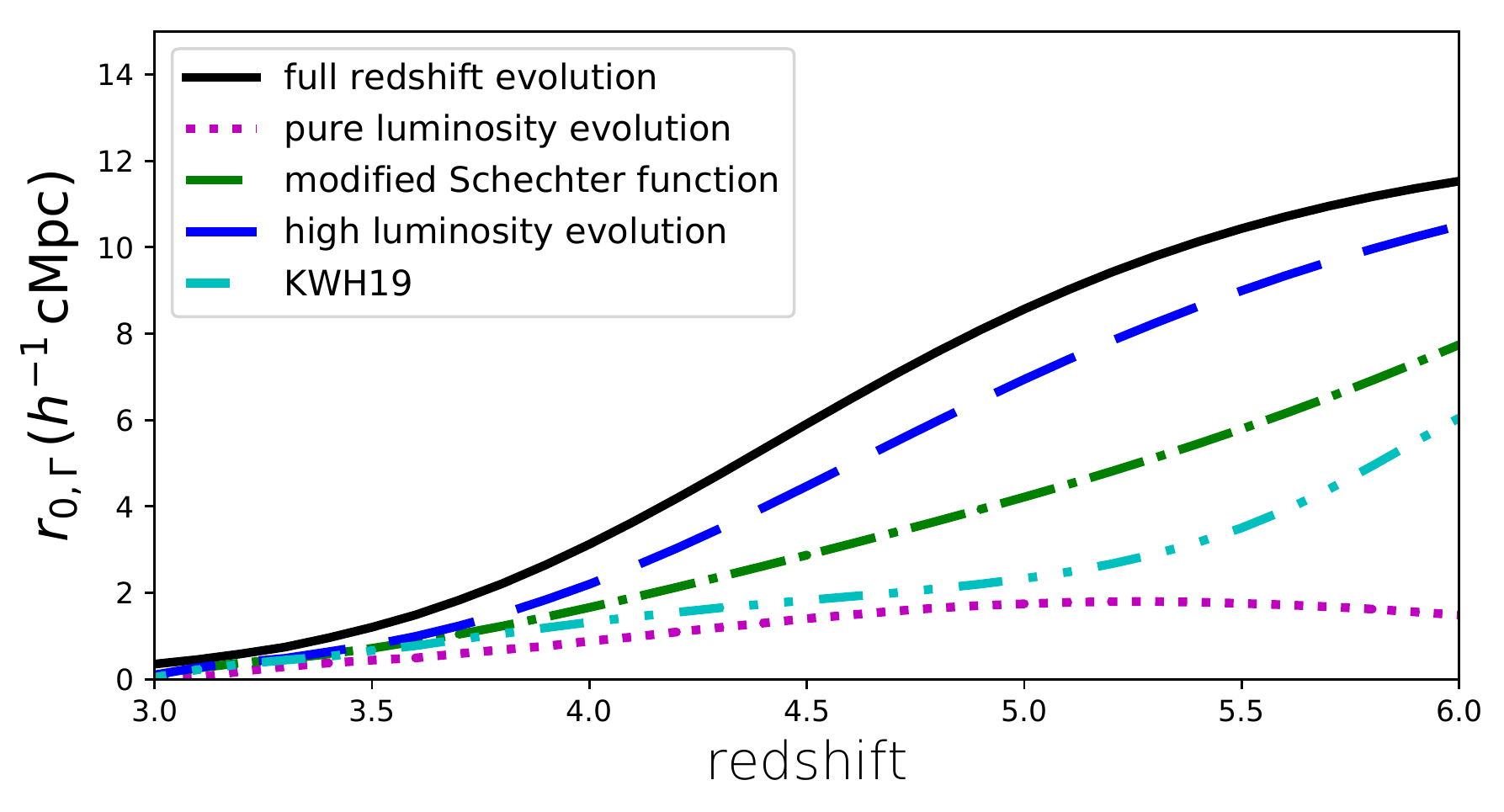}}
\caption{Comoving spatial auto-correlation length for the UV background fluctuations generated by galaxies and QSOs. Four different models for the QSO luminosity function are illustrated:\ a full redshift evolution model (black solid line), a pure luminosity evolution model (magenta dotted line), a modified Schechter luminosity function (green dot-dashed line ), a redshift evolution fit to the high luminosity end of the luminosity function (blue dashed line) and Model 3 from \citet{2019MNRAS.488.1035K} (cyan dot-dot-dashed line).}
\label{fig:LyAXiJr0}
\end{figure}

The UV background correlation length $r_{0, \Gamma}$ may be estimated from the analytic expression for the UV background shot-noise auto-correlation function $\xi^{\rm shot}_{\Gamma\Gamma}(r)$. In the limit of short separations compared with the mean free path $\lambda_{\rm mfp}$ of photoionizing photons, $\xi^{\rm shot}_{\Gamma\Gamma}(r)\rightarrow(\pi/ 16\lambda_{\rm mfp}^3n_{\rm eff})\lambda_{\rm mfp}/r$ \citep{1992MNRAS.258...45Z}. Then
\begin{eqnarray}
  r_{0, \Gamma}&\simeq&\frac{\pi}{16\lambda_{\rm mfp}^3n_{\rm eff}}\lambda_{\rm mfp}\\
  &\simeq&1.3\left(\frac{3\times10^{-6}\,{\rm cMpc}^{-3}}{n_{\rm eff}}\right)\left(\frac{1+z}{5}\right)^{8.8}\,h^{-1}\,{\rm cMpc}\nonumber,
  \label{eq:r0}
\end{eqnarray}
using the mean free path from \citet{2014MNRAS.445.1745W}. For the full redshift evolution model for the QSO luminosity function, the correlation length (based on the full analytic integral), increases rapidly from $\sim0.5h^{-1}\,{\rm cMpc}$ to $\lsim10h^{-1}\,{\rm cMpc}$ over redshifts $z=3$ to 5, as shown in Fig.~\ref{fig:LyAXiJr0}. This would produce a rapid rise in the spread of effective optical depths with redshift, as is observed \citep{2006AJ....132..117F}. The effective source number density, and the consequent UV background correlation length, differ by as much as an order of magnitude between QSO luminosity function models. The correlation strength is weakest for the pure luminosity evolution model, although the correlations are still non-negligible at $z>5$.

The UV background correlation strength, and so the width of the effective optical depth distribution, is sensitive to the uncertain QSO counts, in particular the upper luminosity of the QSOs. The models considered here are based on QSOs with bolometric luminosities $L<10^{15}L_\odot$. The effective number density $n_{\rm eff}$ of sources is very sensitive to the upper limit for the full redshift evolution model, although less so for the pure luminosity evolution model, and insensitive to the upper limit for the modified Schechter function fit \citep{2019MNRAS.482.4777M}.

\begin{figure}
\scalebox{0.55}{\includegraphics{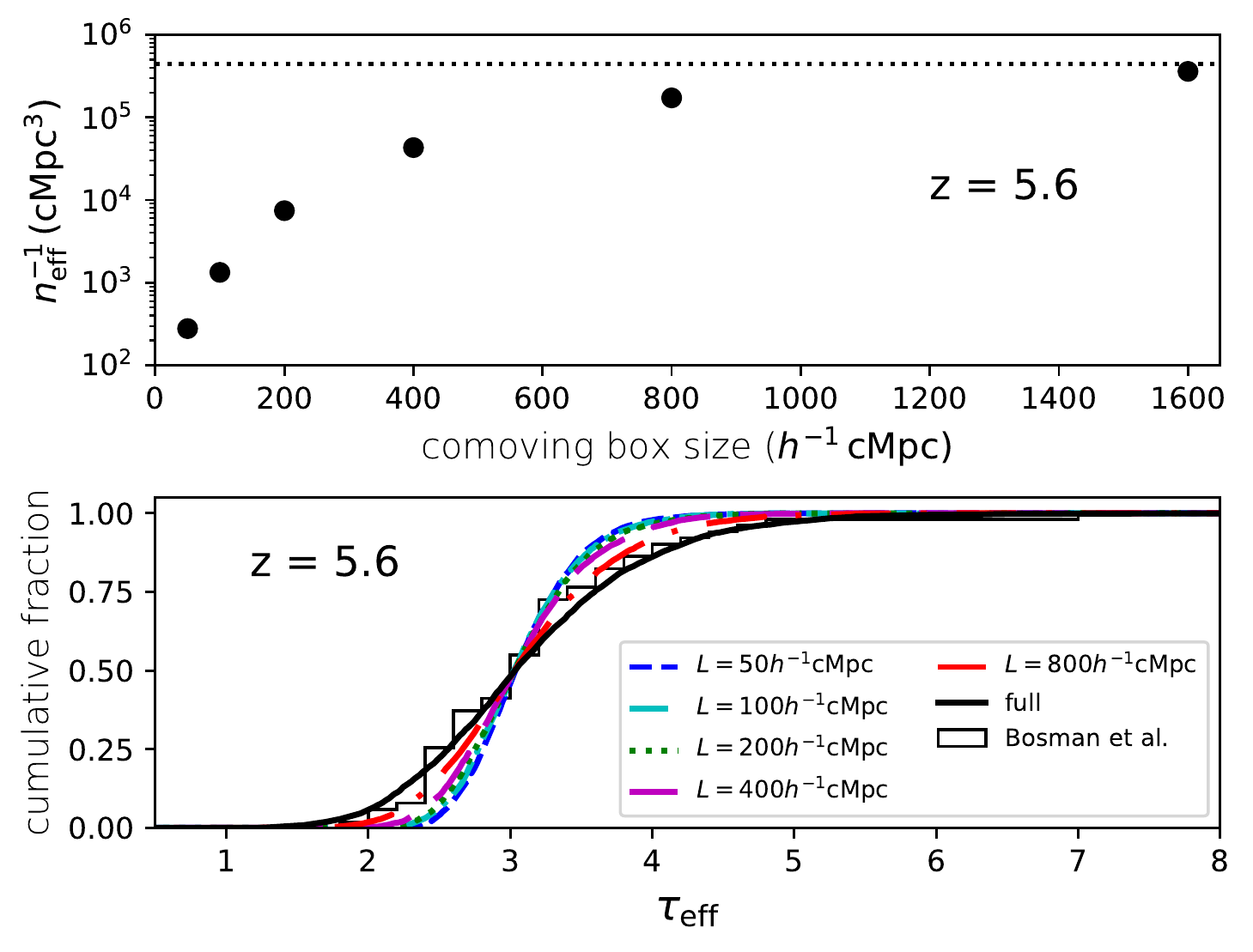}}
\caption{Effect of box size on effective number density of sources (top panel) and the \Lya\ effective optical depth distribution (bottom panel), at $z=5.6$. UV background fluctuations in the steady-state limit are assumed, allowing for both galaxy and QSO sources, for the full redshift evolution QSO luminosity function. The \Lya\ forest model uses a comoving smoothing scale $r_{{\rm f}, \Gamma}=4h^{-1}$~cMpc in the log-normal model for the shot noise in the UV background to compute the cumulative fractions. The histogram shows the data from \citet{2018MNRAS.479.1055B}.}
\label{fig:box_cumtaueff}
\end{figure}

The sensitivity to the most luminous sources may help explain why numerical simulations have been unable to reproduce the broad effective optical depth distributions in scenarios for which reionization completed before $z=6$ without additional assumptions \citep{2015MNRAS.447.3402B, 2015MNRAS.453.2943C}. A too small simulation box size will under-represent the UV background correlations for two reasons:\ (1) insufficient modes captured and (2) inadequate sampling of the QSO luminosity function. As shown in the Appendix, a comoving box size of at least $400h^{-1}\,{\rm cMpc}$ adequately captures the required modes at the redshifts of interest.

Box sizes too small to sample the QSO luminosity function adequately will fail to recover the full strength of the UV background correlations as well. The required box size is estimated by truncating the luminosity function for QSOs at the luminosity above which fewer than a single QSO would be expected in the simulation volume. The resulting values for $n_{\rm eff}^{-1}$ are shown for the full redshift evolution QSO luminosity function model at $z=5.6$, in the upper panel of Fig.~\ref{fig:box_cumtaueff} for a range in box sizes, compared with the actual value (shown as a dotted line). The corresponding effective optical depth distributions are shown in the lower panel. A comoving box size of $800-1600h^{-1}\,{\rm cMpc}$ is required to fully capture the width of the distribution, much exceeding the box sizes of numerical simulations used previously to investigate the effective optical depth distribution.

The lower panel of Fig.~\ref{fig:box_cumtaueff} may also be interpreted as showing the sensitivity of the effective optical depth distribution to the upper limit of QSO luminosities. The most luminous QSOs detected at $z\sim5$ in the analysis of \citet{2007ApJ...654..731H} have an estimated bolometric luminosity of $\sim10^{14}L_\odot$ (corresponding more nearly to the $400h^{-1}\,{\rm cMpc}$ result), so an upper limit of $10^{15}L_\odot$ is an extrapolation of the data (although QSOs more luminous than $10^{15}L_\odot$ are detected at $z<4$). The sampling of high luminosity QSOs, however, is limited by the survey volume.

Other effects suggested to account for the wide spread in effective optical depths without late reionization include a large reduction in the photoionizing mean free path compared with direct measurements from observed QSOs \citep{2016MNRAS.460.1328D, 2018ApJ...863...92B, 2018MNRAS.473..560D}, invoking large temperature fluctuations in the IGM following reionization completing at $z\gsim6$ \citep{2015ApJ...813L..38D} \citep[although][find the effect insufficient in reionization simulations]{2018MNRAS.477.5501K}, or a late reionization scenario, with reionization completing at $z\lsim5.5$, and possibly as late as $z\simeq5.2$ \citep{2019MNRAS.485L..24K, 2019MNRAS.tmp.2682K}.

An alternative solution appeals to an additional population of rare, luminous sources such as QSOs \citep{2015MNRAS.453.2943C, 2017MNRAS.465.3429C}, as in the QSO luminosity function of \citet{2015A&A...578A..83G}, compared with previous estimates. For this model, the QSO and galaxy contributions to the metagalactic emissivity are comparable. Matching to the range of QSO brightnesses $-27<M_{1450}<-22$ in the simulation of \citet{2017MNRAS.465.3429C}, the galaxy contribution must be cut back by about 30 percent so as not to exceed the metagalactic emissivity estimate of \citet{2012ApJ...746..125H}. QSOs then contribute about 30 percent of the total metagalactic source emissivity, compared to 10 percent or less over $5<z<6$ for the other QSO luminosity functions considered here. Along with a somewhat suppressed galaxy contribution, the weight of QSOs increases in the overall effective number density of sources, with values resulting similar to those found for the other QSO luminosity function models (when including the galactic contribution). At $z=5.6$, $n_{\rm eff}\simeq1.2\times10^{-5}\,{\rm cMpc}^{-3}$, just below that found for the PLE model of $n_{\rm eff}\simeq2\times10^{-5}\,{\rm cMpc}^{-3}$. As a consequence, the \Lya\ optical depth distribution will be broadened by large-scale UV background correlations, with results similar to those shown in Fig.~\ref{fig:taueff_z5t6} for the PLE model. Because the luminosity function primarily boosts the number of moderate to low luminosity QSOs compared with the others \citep[see the discussion in][]{2015A&A...578A..83G}, the effective number density of sources is not very sensitive to the bright end of the luminosity function. The box size of $500h^{-1}\,{\rm cMpc}$ in \citet{2017MNRAS.465.3429C} is sufficiently large to capture 65 percent of the strength of the UV background correlations. It has been noted, however, that the addition of such a large number of QSOs risks under-predicting the measured \HeII\ optical depths and overheating the IGM at lower redshifts \citep{ 2017MNRAS.468.4691D, 2019MNRAS.483.5301G, 2019MNRAS.485...47P}. The QSO number counts have also since come down somewhat \citep{2019ApJ...884...19G}, with reduced effective number densities when combined with galaxies (see end of Sec.~\ref{subsec:uvbgmodel} above).

The models presented in this paper rely exclusively on conservative estimates of the galaxy and QSO populations, IGM temperatures and the intergalactic mean free path of ionizing photons. The only \lq tuneable\rq\ parameter that has much influence is the mean UV background flux required to match the median measured optical depths, and the values obtained are comparable to previous estimates \citep{2012ApJ...746..125H, 2019MNRAS.485...47P}.

Other observational tests of the model include the distribution of \Lyb\ (and higher order) optical depths, the statistics of transmission spikes in the \Lya\ forest spectra \citep{2008MNRAS.386..359G, 2019ApJ...876...31G}, and the clustering statistics of \Lya\ emitters, as modulated by foreground IGM absorption \citep{2019MNRAS.tmp.2682K}. These topics are deferred to future work.

\section{Conclusions}
\label{sec:conclusions}

The broadening distribution of \Lya\ effective optical depths over wide patches at $z>5$  has posed a challenge to models of the IGM. The distributions are broader than expected for a uniform UV background. Various suggestions have been made to account for the wide range in values, including higher numbers of QSO sources at high redshifts than discovered in previous surveys, shorter mean free paths for ionizing photons than inferred directly from QSO spectra, late reionization ($z<6$), with large patchy mean free paths and temperature fluctuations remaining, or that the reionization process itself may be incomplete until as late as $z\simeq5.2$.

Using semi-analytic models for the UV background fluctuations and for the \Lya\ forest, it is suggested here instead that the broad distributions are consistent with conservative estimates of the galaxy and QSO counts, and in fact are expected as a consequence of the UV background fluctuations produced by shot noise from the sources. Although QSOs contribute only $\sim10$ percent of the mean UV background at $z>5$, they dominate the shot noise, resulting in large-scale spatial correlations in the UV background with a comoving correlation length of order $1-10h^{-1}\,{\rm cMpc}$, depending on QSO luminosity function and redshift. These correlations may account for the wide range in effective optical depths measured over comoving spatial intervals of $50h^{-1}\,{\rm cMpc}$.

Because of the approximate nature of a semi-analytic approach, full 3D simulations are required for detailed comparison with the data. Capturing the full extent of the spatial correlations in the UV background, however, is numerically challenging, requiring simulation volumes of at least $400h^{-1}\,{\rm cMpc}$ on a side just to include the required wavemodes. For a single simulation to recover the full contribution of QSO sources to the UV background fluctuations places an even greater demand of box sizes of $\sim1$~cGpc to adequately sample the QSO luminosity function. Because of the comparatively short mean free path of ionising photons, a $400h^{-1}\,{\rm cMpc}$ box may be adequate if results are averaged over sufficient random realizations ($\sim10-100$) of the QSO population.

The conclusions in this paper do not exclude the possibility of late reionization. The distribution in the mean optical depths is highly sensitive to the QSO luminosity function and its evolution, especially to the numbers of the most luminous QSOs. Simulations invoking late reionization none the less must also adequately sample the QSO luminosity function to ensure they recover the UV background fluctuations induced. Estimates here suggest a finite lifetime for the QSOs will also be a factor, but is not as important as the QSO luminosity function itself. Precision estimates, however, including finite QSO lifetimes would place the further demand on the simulations of including light cone effects. Until such simulations are performed, it appears still an open question as to whether or not the broad high redshift \Lya\ effective optical depth distributions may be attributed to late reionization, modifications of the QSO or galaxy luminosity functions, modifications to the structure of the IGM, or are primarly a consequence of the expected QSO and galaxy shot-noise induced UV background fluctuations.

\section*{Acknowledgements}
AM thanks an anonymous referee for a careful reading of the manuscript and suggestions that improved the clarity of the presentation. AM acknowledges support from the UK Science and Technology Facilities Council, Consolidate Grant ST/R000972/1.

\bibliographystyle{mn2e-eprint}
\bibliography{ms}

\appendix

\section{Extended log-normal model for the \Lya\ forest}
\label{sec:logn_ap}

The density, velocity and photoionization fields used to construct the \Lya\ forest spectra depend on 1D projected power (and cross-power) spectra. In terms of the 3D (cross-) power spectrum $P_{ij}(k)$ between objects $i$ and $j$, the 1D power spectrum projected along the $z$ direction is
\begin{equation}
  P^{\rm 1D}_{ij}(k_z,z)=\frac{1}{2\pi}\int_{k_z}^\infty\,dk k P_{ij}(k,z).
  \label{eq:P1D}
\end{equation}
The Fourier components of the baryon density and peculiar velocity perturbations and density-dependent UV background perturbation are expressed, respectively, as
\begin{eqnarray}
  \delta_b(k_z,z) &=& u_1 + u_2 + u_3,\nonumber\\
  \frac{w(k_z,z)}{k_z} &=&au_2 + bu_3,\nonumber\\
\noindent{\rm and}\nonumber\\
  \delta_{\Gamma, b}(k_z,z) &=&cu_3,
\label{eq:deltas}
  \end{eqnarray}
where $u_1$, $u_2$ and $u_3$ are independent (complex) Gaussian random deviates. The 1D cross power spectra are reproduced by setting
\begin{eqnarray}
  a &=& \frac{P^{\rm 1D}_{ww}(k_z,z)P^{\rm 1D}_{\Gamma\Gamma}(k_z,z)-\left[P^{\rm 1D}_{w\Gamma}(k_z,z)\right]^2}{P^{\rm 1D}_{bw}(k_z,z)P^{\rm 1D}_{\Gamma\Gamma}(k_z,z)-P^{\rm 1D}_{w\Gamma}(k_z,z)P^{\rm 1D}_{b\Gamma}(k_z,z)},\nonumber\\
  b&=& \frac{P^{\rm 1D}_{w\Gamma}(k_z,z)}{P^{\rm 1D}_{b\Gamma}(k_z,z)},\nonumber\\
\noindent{\rm and}\nonumber\\
  c&=& \frac{P^{\rm 1D}_{\Gamma\Gamma}(k_z,z)}{P^{\rm 1D}_{b\Gamma}(k_z,z)}.
\label{eq:abc}
\end{eqnarray}
Here, the subscript $w$ refers to $w(k_z,z)/k_z$.

The values for $u_1$, $u_2$ and $u_3$ are chosen using the polar decomposition $u_i=|u_i|e^{i\phi_i}$, where $\phi_i$ is drawn from a uniform random distribution over $[0,2\pi]$ and $|u_i|=\bkts{P_i(k_z,z)/2}^{1/2}\sqrt{-2\log\mc{X}}$, where $\mc{X}$ is a uniform random deviate over $[0,1]$, and $P_i(k)$ is the power spectrum for $u_i$. The power spectra are given by
\begin{eqnarray}
P_1(k_z,z) &=& P^{\rm 1D}_{bb}(k_z,z) - P_2(k_z,z) - P_3(k_z,z),\nonumber\\
\noindent{\rm where}\nonumber\\
P_2(k_z,z) &=& \frac{\bkts{P^{\rm 1D}_{bw}(k_z,z)P^{\rm 1D}_{\Gamma\Gamma}(k_z,z)-P^{\rm 1D}_{b\Gamma}(k_z,z)P^{\rm 1D}_{w\Gamma}(k_z,z)}^2}{P^{\rm 1D}_{\Gamma\Gamma}(k_z,z)\bkts{P^{\rm 1D}_{ww}(k_z,z)P^{\rm 1D}_{\Gamma\Gamma}(k_z,z)-\bkt{P^{\rm 1D}_{w\Gamma}(k_z,z)}^2}}\nonumber\\
\noindent{\rm and}\nonumber\\
P_3(k_z,z) &=& \frac{\bkts{P^{\rm 1D}_{b\Gamma}(k_z,z)}^2}{P^{\rm 1D}_{\Gamma\Gamma}(k_z,z)}.
\end{eqnarray}
In the limit $\delta_\Gamma\rightarrow0$, $P_3=0$, $P_2\rightarrow (P^{\rm 1D}_{bw})^2/ P^{\rm 1D}_{ww}$, where
\begin{eqnarray}
  P^{\rm 1D}_{bw}(k_z,z)&=&\frac{1}{2\pi}\int_{k_z}^\infty\,dk k^{-1}P_{bb}(k,z),\nonumber\\
  P^{\rm 1D}_{ww}(k_z,z)&=&\frac{1}{2\pi}\int_{k_z}^\infty\,dk k^{-3}P_{bb}(k,z),
\end{eqnarray}
$b=c=0$, and $a\rightarrow P^{\rm 1D}_{ww}(k_z,z)/ P^{\rm 1D}_{bw}(k_z,z)$,
recovering the result of \citet{1997ApJ...479..523B}.

The shot-noise contribution $\delta_{\Gamma, {\rm sn}}$ is approximated as a (complex) Gaussian random deviate using the filtered power spectrum
\begin{equation}
P^{\rm 1D}_{\Gamma\Gamma, {\rm sn}}(k_z,z)=\frac{1}{2\pi}\int_{k_z}^\infty\,dk \frac{k}{\bkts{1+(r_{f, \Gamma}k)^2}^2} P_{\Gamma\Gamma, {\rm sn}}(k,z),
\end{equation}
where a Lorentzian filter of width $r_{f, \Gamma}$ has been adopted, and $P_{\Gamma\Gamma, {\rm sn}}(k,z)$ is the shot-noise power spectrum \citep{2019MNRAS.482.4777M}. The filtering is required because the 1-point fluctuations in the photoionization rate induced by the shot noise are non-linear; in fact the variance at a single spatial point diverges \citep{MW03}. Three approaches are used to approximate the shot-noise contribution: (1)\ broadly filtering the fluctuations to obtain linear fluctuations, (2)\ adopting the 1-point flux distribution but including spatial correlations in a gaussian approximation, and (3)\ adopting a log-normal distribution for the fluctuations, following Eq.~(\ref{eq:Gammalogn}). Each of these is discussed in turn. All the tests here use the steady-state UV background power spectrum predicted in linear theory with shot noise \citep{2019MNRAS.482.4777M}, for galaxy and QSO sources, assuming the full redshift evolution model for the QSO luminosity function.

\begin{figure*}
\includegraphics{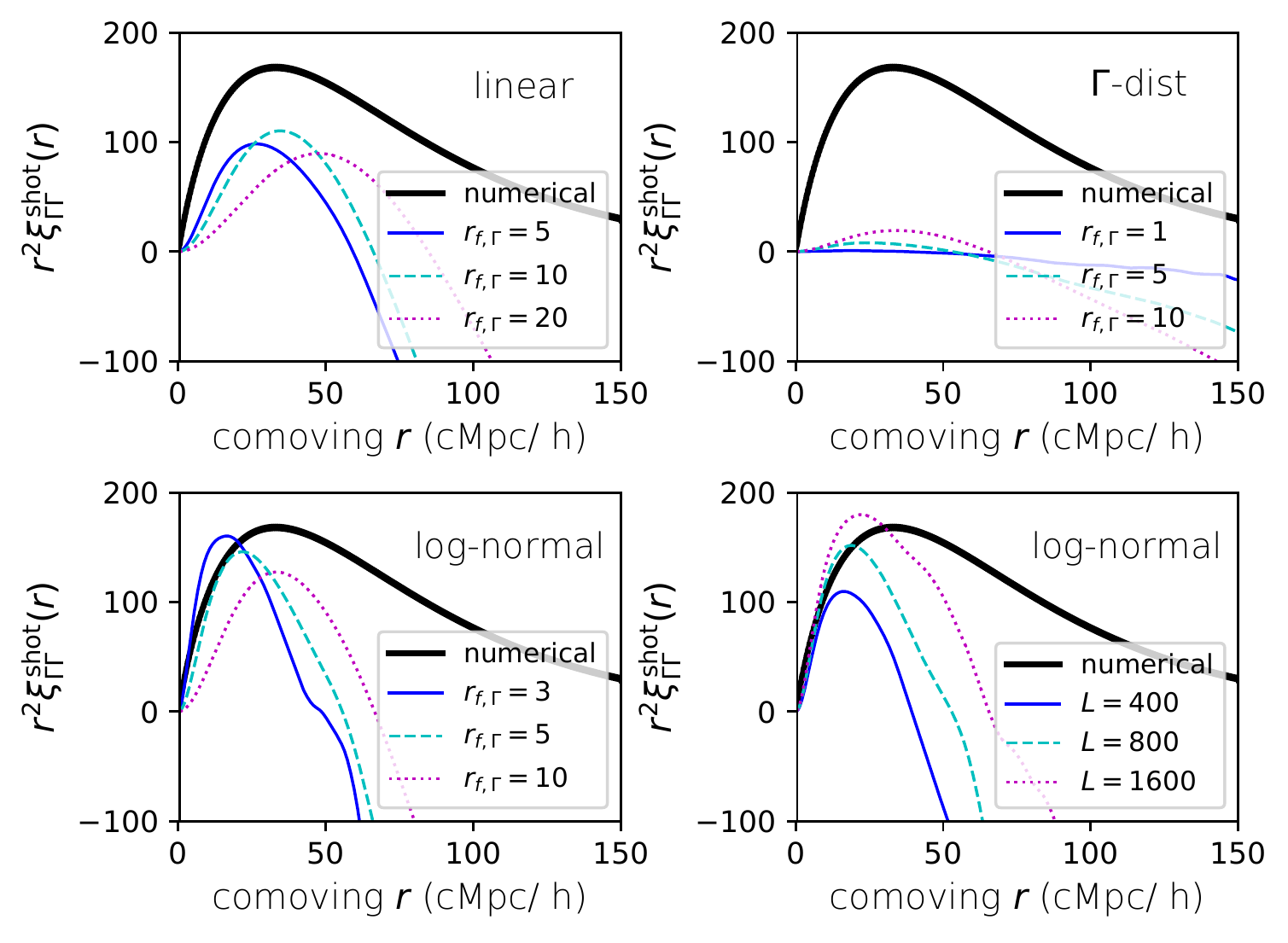}
\caption{Spatial auto-correlation function of UV background fluctuations, as a function of comoving separation. The heavy solid (black) line shows the numerical prediction. The indicated filter lengths $r_{{\rm f}, \Gamma}$ and line-of-sight spectrum length $L$ are in units of comoving $h^{-1}\,{\rm cMpc}$. Shown at redshift $z=5.6$. (Top left panel):\ Linear UV background fluctuation model. (Top right panel): 1-point shot-noise $\Gamma$ distribution. (Bottom left panel):\ Log-normal UV background fluctuation model. (Bottom right panel):\ Convergence test of log-normal UV background fluctuation model for different spectrum line-of-sight lengths, for $r_{{\rm f}, \Gamma}=4h^{-1}\,{\rm cMpc}$.}
\label{fig:XiGamma}
\end{figure*}

\begin{figure}
\scalebox{0.55}{\includegraphics{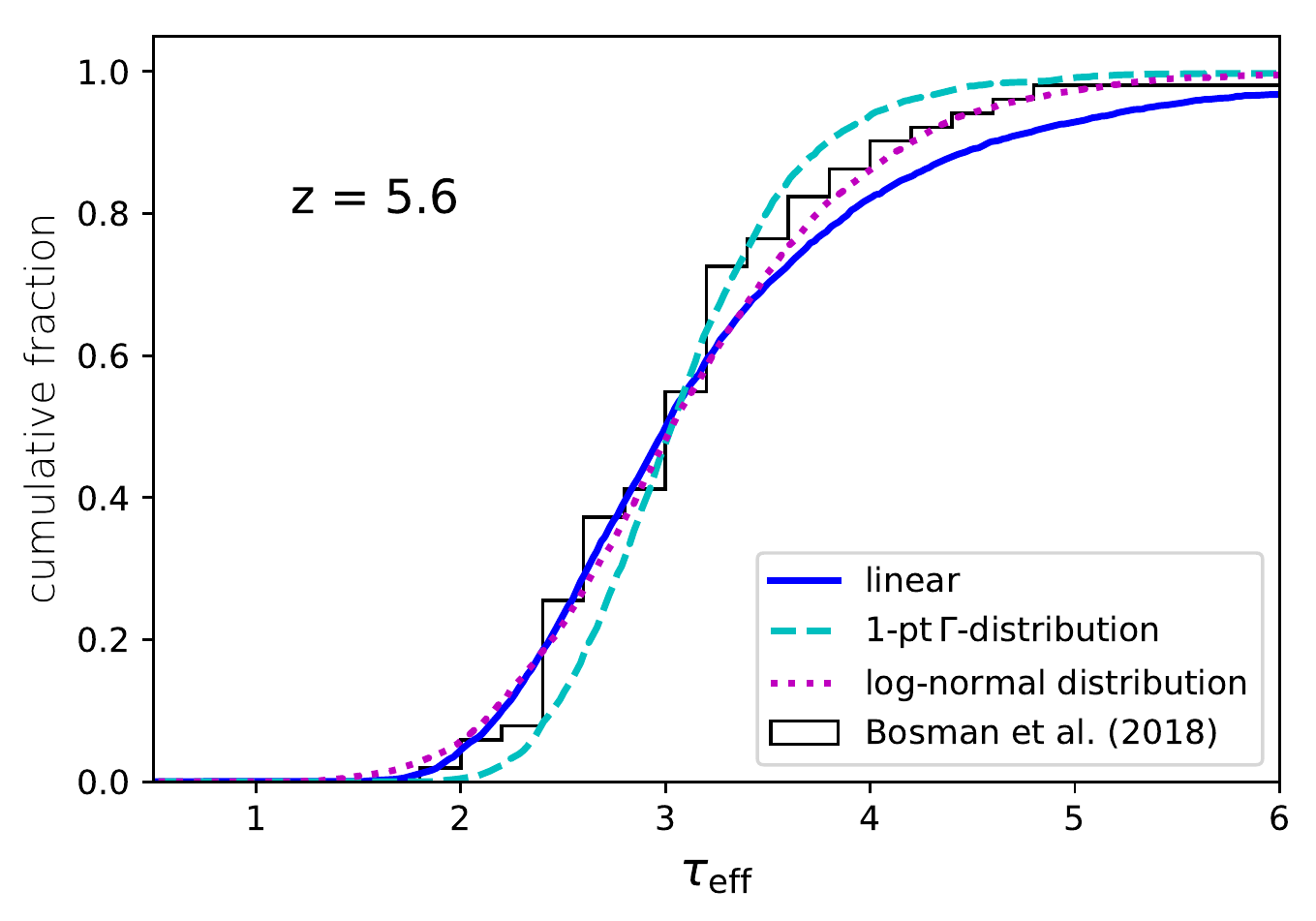}}
\caption{Cumulative distribution of mean optical depths, averaged over (comoving) segments $50h^{-1}\,{\rm cMpc}$ long. The histogram shows the data from \citet{2018MNRAS.479.1055B}. From left to right at their bases, the smooth curves show the predicted distributions from the log-normal UV background fluctuation model ($r_{{\rm f}, \Gamma}=4h^{-1}\,{\rm cMpc}$), the linear UV background fluctuation model ($r_{{\rm f}, \Gamma}=5h^{-1}\,{\rm cMpc}$) and a model based on the 1-pt $\Gamma$ distribution ($r_{{\rm f}, \Gamma}=10h^{-1}\,{\rm cMpc}$).}
\label{fig:cumtaueff_models}
\end{figure}

\begin{figure}
\scalebox{0.55}{\includegraphics{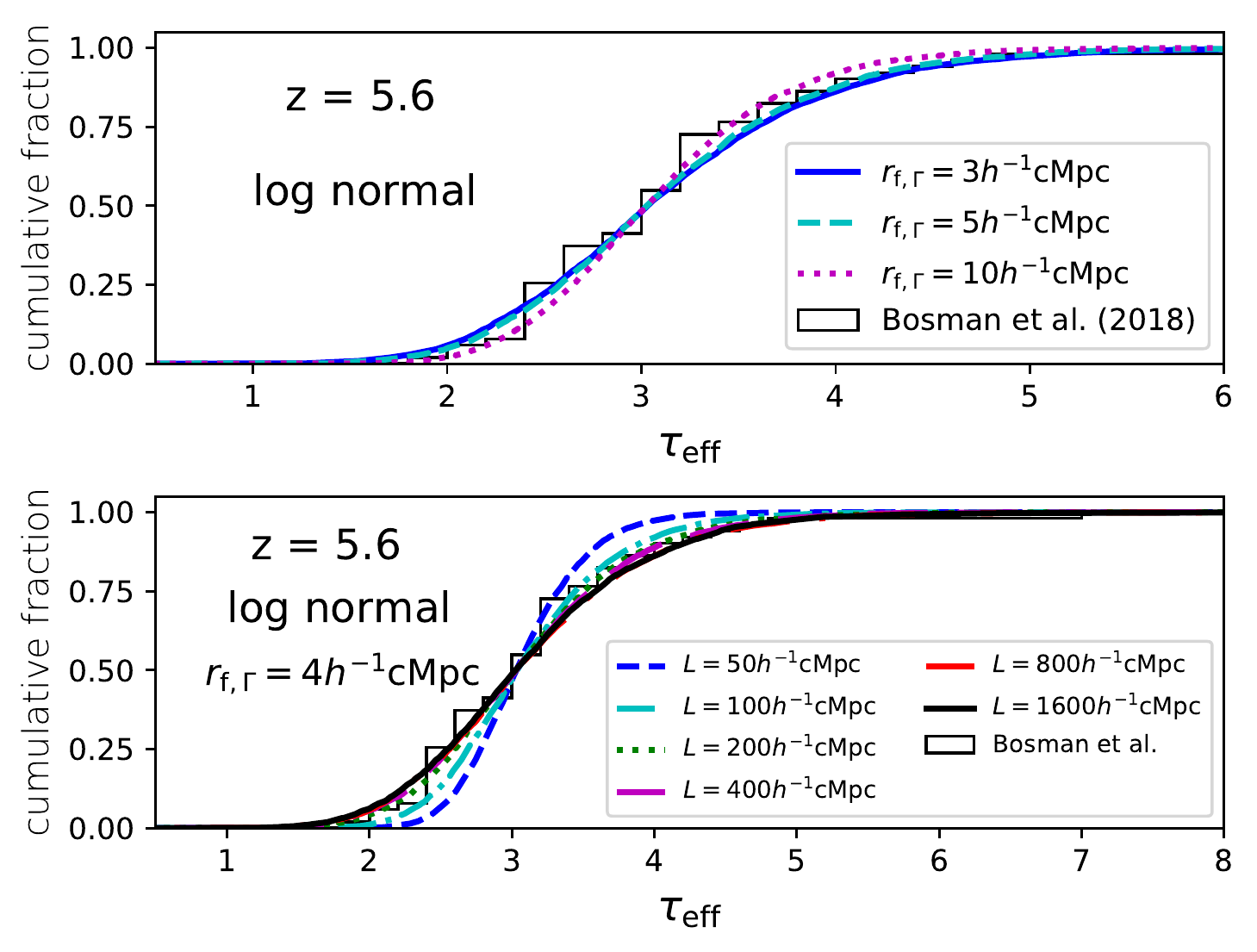}}
\caption{Cumulative distribution of mean optical depths, averaged over (comoving) segments $50h^{-1}\,{\rm cMpc}$ long. The histrograms show the data from \citet{2018MNRAS.479.1055B}. The smooth curves are predictions from the log-normal UV background fluctuation model, for varying filter scales $r_{{\rm f}, \Gamma}$ (top panel) or line-of-sight spectrum length $L$ (bottom panel). The length units are comoving. From left to right at their bases, the curves are for increasing $r_{{\rm f}, \Gamma}$ or decreasing $L$.}
\label{fig:cumtaueff_rfG_L_conv}
\end{figure}

Using smoothed linear fluctuations alone is problematic because over-smoothing fails to reproduce the spatial correlations in the UV background, as shown in the upper left panel of Fig.~\ref{fig:XiGamma}, while reducing the smoothing results in too large negative excursions in the fluctuations. The negative excursions must be truncated to ensure $\delta_{\Gamma}>-1$ (to prevent negative photoionization rates), artificially creating neutral patches and an extended tail of large values in the effective optical depth distribution, as shown in Fig.~\ref{fig:cumtaueff_models}.

As an alternative, the 1-point distribution in the UV background fluctuations is adopted \citep{MW03}, mapping the frequency distributions between the 1-point fluctuations to random gaussian fluctuations chosen from the shot-noise UV background power spectrum. The motivation is to attempt to retain the expected range in non-linear fluctuations whilst realizing the shot-noise induced spatial correlations in the UV background. Unfortunately the non-linear map between the gaussian distribution and the 1-point distribution suppresses the correlations, as shown in the upper right panel of Fig.~\ref{fig:XiGamma}. The resulting effective optical depth distribution is essentially unchanged from the uniform UV background case (Fig.~\ref{fig:cumtaueff_models}).

Instead a log-normal distribution is adopted. This has the advantages of recovering the spatial correlations in the UV background where they are strong, as shown in the bottom left panel of Fig.~\ref{fig:XiGamma}, relative insensitivity of the predicted effective optical depth distribution to the smoothing scale, as shown in the top panel of Fig.~\ref{fig:cumtaueff_rfG_L_conv}, and qualitatively matching the expected sharp cut-off in low excursions of the UV background fluctuations and broad tail of non-linear high excursions expected as the smoothing length is decreased \citep{MW03}. It also converges to the linear limit for sufficently large smoothing.

Converging on spatial correlations in a simulation volume is computationally demanding. To estimate the required box sizes, spectra are generated for lines of sight of varying lengths $L$, using the full shot noise power spectrum. The length scale over which the log-normal model recovers the shot-noise induced spatial correlations in the UV background increases with $L$, as shown in the lower right panel of Fig.~\ref{fig:XiGamma}. As shown in the lower panel of Fig.~\ref{fig:cumtaueff_rfG_L_conv}, a box size of $400h^{-1}$~cMpc or larger is required to converge on the effective optical depth distribution. This is large enough to recover moderate to large spatial correlations in the UV background ($\xi_{\Gamma\Gamma}^{\rm shot}>0.5$).

\label{lastpage}
\end{document}